\documentclass[12pt]{elsarticle}
\usepackage{graphicx}
\usepackage{dcolumn}% Align table columns on decimal point
\usepackage{bm}% bold math
\newcommand{\Sec}[1]{Section~\ref{sec:#1}}

\newcommand{\Eq}[1]{Eq.\ (\ref{eq:#1})}
\newcommand{\Eqs}[2]{Eqs~(\ref{eq:#1})~and~(\ref{eq:#2})}

\usepackage{latexsym}

\begin{document}

%\ead{wayne.arter@ccfe.ac.uk}

\title{An Anti-Perfect Dynamo Result}
\author{Wayne Arter}
\address{EURATOM/CCFE Fusion Association, Culham Science Centre, Abingdon, UK. OX14 3DB
\\wayne.arter@ccfe.ac.uk,
+44 1235 466498 (Phone),
+44 1235 466435 (Fax)
}
\date{\today}

\begin{abstract}
It is shown that if the flow vector is a coordinate basis vector, then perfect
dynamo action is not possible, regardless of whether the steady
flow is compressible. Criteria determining the basis vector property
are found to be expressible in terms of Lie derivatives, and construction
of basis vector sets is straightforward.\\
%wayne.arter@ccfe.ac.uk \\
%+44 1235 466498 (Phone) \\
%+44 1235 466435 (Fax)
\end{abstract}

%\pacs{52.35.-g, 47.55.P-, 02.30.Oz}

\maketitle

\section{Introduction}\label{sec:intro}
%?Compressible hydrodynamics may be formulated in terms of
%?a Lie derivative of a vector by introducing the potential vorticity,
%?the vorticity (ie.\ curl of the velocity field) divided by the mass density,
%?a result which goes back to Helmholtz~\cite[\S\,146]{lamb}.
The magnetic induction equation for compressible flow may
be formulated in terms of
a Lie derivative of a vector by introducing the field defined as the
the magnetic field~${\bf B}$ divided by the mass density.
This  result is originally due to Walen~\cite[\S\,4-2]{boydsanderson},
and is re-derived in component form, in
for example Chandrasekhar~\cite[\S\,38(b)(i)]{H2S} and Roberts~\cite[\S\,2.3(c)]{roberts}:
\begin{equation}\label{eq:induc}
\partial\tilde{\bf B}/\partial t=\tilde{\bf B}.\nabla{\bf u}-{\bf u}.\nabla\tilde{\bf B}=
\mathcal{L}_{\bf u}(\tilde{\bf B})
\end{equation}
where $\mathcal{L}_{\bf u}$ is the Lie derivative with respect to the
flow field~${\bf u}$, $\tilde{\bf B}={\bf B}/\rho$ and $\rho$ is mass density.
The Lie derivative and related geometrical concepts are
explained further in \Sec{math}.
%?Indeed all the time-dependent equations of incompressible magnetohydrodynamics
%?have been put into a Lie derivative form~\cite{Wa13a}.

This work focusses on the perfect dynamo problem rather than the fast
dynamo, see \Sec{dyna} for a discussion as to how they are interrelated.
The perfect problem may
be posed in terms of solutions of~\Eq{induc}, namely the flow~${\bf u}$ is
said to be a perfect dynamo action if volume-integrated
absolute fluxes of~${\bf B}$ grow exponentially in time
for some seed field~\cite{gilbert}.

At the root of the current work
is the simple observation~\cite[Ex.\,3.3]{schutz} that if the vector~${\bf u}$ in
the Lie derivative~$\mathcal{L}_{\bf u}(\tilde{\bf B})$ may be identified as a basis vector~${\bf e}_3$
in a coordinate basis, then \Eq{induc} reduces to pure advection, viz.
\begin{equation}\label{eq:adv}
\partial\tilde{B}^i/\partial t= \partial_3 \tilde{B}^i 
\end{equation}
where $\tilde{B}^i$ are the contravariant components of~$\tilde{\bf B}$
and $\partial_3$ is a shorthand for $\partial/\partial \bar{x}^3$,
the derivative with respect to the coordinate corresponding to~${\bf u}$.
(See the start of \Sec{steady} for
proof of the reduction \Eq{adv}, which is  almost immediate given the relevant
geometrical concepts from \Sec{math}.)
\Eq{adv} immediately implies that $\tilde{B}^i$ is conserved following the
flow and, as discussed in \Sec{steady}, rules out perfect dynamo action
for all flows in the aforementioned class.

Hence, this generalises the known anti-dynamo results that require
invariance %(usually of~${\bf B}$)
in one Cartesian coordinate, see \cite[Chap.\,V]{arnoldkhesin},\cite{Gi03Dyna}. 
It is to be compared to the result that a perfect dynamo action
requires a finite amount of topological
entropy~\cite{Kl95Rigo}.
This property is not obvious to compute
whereas the properties of coordinate basis vectors are simply
described, and the basis vector property is easy to test in a
set of three vectors. Remarkably, the principal test
relies on establishing a property
which involves Lie brackets (equivalent entities
to Lie derivatives of vectors), namely that the Lie brackets of
basis vectors vanish.

The result established herein is in some respects
not as general as that of~\cite{Kl95Rigo} viewed as an anti-dynamo
theorem for flows with zero topological entropy, but in other respects,
it represents an important extension
in that there is no smoothness requirement on the initial ${\bf B}$-field,
for example.
%Further, ref~\cite{Kl95Rigo}
%considers only flows described by diffeomorphisms, ie. coordinate basis
%mappings in the language of the present paper, which the results herein
%do not require in every case.

\section{Mathematics}\label{sec:math}
\subsection{Dynamo Definitions}\label{sec:dyna}
Fast dynamo action is defined in the context of the `classic'
version of the magnetic induction
equation with resistivity~$\epsilon$, viz.
\begin{equation}\label{eq:induceps}
\partial{\bf B}/\partial t=\nabla\times({\bf u}\times{\bf B}) +\epsilon \nabla^2{\bf B}
\end{equation}
If the magnetic energy of solutions to \Eq{induceps}
grows exponentially with positive growth rate
in the limit $\epsilon \rightarrow 0$ (for some initial seed field),
then the flow~${\bf u}({\bf x})$ is said to be a fast dynamo.
The results presented herein concern the case $\epsilon=0$,
${\bf u}({\bf x})$
is a perfect dynamo if fluxes of magnetic field grow exponentially, strictly
that the following inequality be satisfied
\begin{equation}\label{eq:aflux}
\mbox{Lim}_{t\rightarrow\infty}\frac{1}{t}\log\left(\int_V|{\bf B}({\bf x},t)|dV\right)> 0
\end{equation}

The two mathematical models are physically quite different in that nonzero~$\epsilon$
is more realistic as even tiny amounts of dissipation prevent the appearance
of field singularities. The two are clearly related however.
%for it is reasonable to conjecture that since
%diffusion only acts to weaken
%dynamo action, if a flow cannot be a perfect dynamo, it cannot be a fast dynamo either.
For example, ref~\cite{Kl95Rigo} establishes the result that the
topological entropy criterion for perfect dynamo action also ensures fast dynamo action.

\subsection{Geometry}\label{sec:geom}
This section collates relevant information from textbooks such as refs\cite{fecko,schutz,frankel},
see ref~\cite{Wa13c} for pointers to specific sections of these works.

A set of three vectors~$\{{\bf e}_i,\,\,i=1,2,3\}$, forms a basis in 3-D provided the vectors
are linearly independent at each point. The vectors are said to form a
coordinate basis if each may be parameterised by~$\bar{x}^i$ such that the
$\bar{x}^i$ may be used as a set of coordinates. Thus a coordinate basis is
determined by a mapping from parameter
space to real space~${\bf x}(\bar{x}^1,\bar{x}^2,\bar{x}^3)$ such that
\begin{equation}\label{eq:coordv}
{\bf e}_i=\partial{\bf x}/\partial\bar{x}^i,\;\;\;i=1,2,3
\end{equation}
form a basis. At this point, it is convenient to define the metric tensor
$g_{ij}={\bf e}_i.{\bf e}_j$, and the local volume or ``volume element"
$\sqrt{g}=\sqrt{\det(g_{ij})}$.

Now in any physically reasonable 3-D coordinate system, there is the
remarkable result that the Lie derivative of a vector may be written
\begin{equation}\label{eq:lieder}
{\bf\mathcal{L}}_{\bf u}({\bf v})^i= v^k\frac{\partial u^i}{\partial x^k}-u^k\frac{\partial v^i}{\partial x^k}
\end{equation}
Hence the Lie bracket notation as an equivalent for the Lie derivative
\begin{equation}
\mathcal{L}_{\bf u}({\bf v})=[{\bf u},{\bf v}]
\end{equation}
so that the two vectors~${\bf u}$ and~${\bf v}$  appear on an equal
footing. Adopting this notation, it may be shown
the~${\bf e}_i$ defined by \Eq{coordv} satisfy
\begin{equation}\label{eq:coordb}
[{\bf e}_i,{\bf e}_j]={\bf 0},\;\;\; \forall i,\;j
\end{equation}
and that if \Eq{coordb} holds, linearly independent vectors~$\{{\bf e}_i,\,\,i=1,2,3\}$
form a coordinate basis. (Strictly speaking, this last statement requires
the Poincare lemma, which applies in toroidal geometry
only if the toroidal angles are allowed free range, ie.\ not restricted
to $[0,2\pi]$, a caveat discussed in more detail in ref~\cite[\S\,II.A]{Wa13c}.)
Note that $\{\lambda({\bf x}){\bf e}_i\}$ is not in general a coordinate basis
unless $\lambda=\mbox{const}\neq 0$, for example
\begin{equation}\label{eq:lielam}
[\lambda{\bf e}_1,\lambda{\bf e}_2]
=\lambda^2[{\bf e}_1,{\bf e}_2]+ \lambda \left(
{\bf e}_2 \partial\lambda /\partial \bar{x}^1- {\bf e}_1 \partial \lambda /\partial \bar{x}^2
\right)
\end{equation}

To establish anti-dynamo action, it is necessary to show that
${\bf u}$ may be expressed as a coordinate basis vector. This is equivalent to
showing that there exists a
mapping~${\bf x}(\bar{x}^1,\bar{x}^2,\bar{x}^3)$ such that
\begin{equation}\label{eq:dxdx}
{\bf u} = \partial {\bf x} / \partial \bar{x}^3
\end{equation}
for some coordinates~$\bar{x}^i$, and ${\bf u} \neq {\bf 0}$.
For example, $\bar{x}^3$ could be a function of arc length
along a streamline. To complete the coordinate system, streamline labels
in a plane normal to~${\bf u}$ (cf.\ Clebsch variables) could be
used. Note that the possibility of such a coordinate basis system
depends on the flow's having a relatively simple topology, see also
Roberts~\cite[\S\,2.5(b)]{roberts} for discussion.
Regardless, the minimal requirement for the other two vectors
say~${\bf e}'_i,\;\;i=1,2$ in the frame is that $[{\bf e}_3,{\bf e}'_i]=0$.
%?Note that if there is a  mapping~${\bf x}(\bar{x}^1,\bar{x}^2,\bar{x}^3)$,
%?then \Eq{dxdx} implies
%?\begin{equation}
%?{\bf u} = \sqrt{g} \nabla \bar{x}^1 \times \nabla \bar{x}^2
%?\end{equation}
%?Compare the Clebsch representation of a vector field.

Limiting the applicability of many of the results which follow is
the Poincare-Hopf theorem relating the number of zeroes of a vector
field to the topology of the  compact manifold on which it is
defined (which gives the `hairy-ball' theorem in the case of spherical surfaces).
For a vector field to form part of a basis, it is obviously necessary
that it be non-zero everywhere, implying that the only compact coordinate systems
are to be found in a toroidal geometry. Thus, apart from
these toroidal systems, all the dynamo results in this section have their
application restricted to unbounded flows.

\section{Detailed Analysis}\label{sec:detail}
\subsection{Steady flows}\label{sec:steady}
The anti-dynamo result outlined in
\Sec{intro} requires further discussion. Using the results
of the previous section, \Eq{adv} is derived as follows.
Writing $\tilde{\bf B}= \tilde{B}^i {\bf e}_i$, and substituting
in \Eq{induc}, using \Eq{coordb}, gives
\begin{equation}\label{eq:inducv}
\partial\tilde{\bf B}/\partial t={\bf e}_i \partial\tilde{B}^i/\partial \bar{x}^3
\end{equation}
Taking components gives \Eq{adv}.
Although the $\tilde{B}^i$ are simply advected from place to place,
the physical magnetic field
is given by $B^i {\bf e}_i$ (summation convention applies),
so will change over time according as the
basis vector~${\bf e}_i$ changes with position. Similarly, since it is
actually  $\tilde{\bf B}={\bf B}/\rho$ which is conserved, there is a further factor
due to the change in the volume element~$\sqrt{g}$. This may be seen from
the  equation for conservation
of mass $m = \rho \sqrt{g}$, which is also expressible in terms of a Lie derivative
and so reduces to
\begin{equation}
\partial m/\partial t= \partial_3 m
\end{equation}

However, for a coordinate system without singularities, derived
geometric quantities such as~${\bf e}_i$ and~$\sqrt{g}$ are well-behaved,
in particular $\sqrt{g}\neq 0$. Thus although  the physical~${\bf B}$
changes, an exponential increase without limit is ruled out.

The absence of singularities is important as it rules  from
consideration such situations as a purely radial inflow: either
the flow continues to the coordinate singularity or it is
stopped on some surface. The latter condition obviously
requires ${\bf u}$ to vanish, and at such points, it cannot
be part of a basis.

\subsection{Time dependent flows}
In the case of time dependent flows, 
an anti-dynamo result might be established for a ${\bf u}$ 
which is one of a set of time varying coordinate basis vectors,
only if further conditions are placed on the basis variation.
Compressibility also changes the mass conservation equation because
$m$ only evolves as a Lie derivative if $\sqrt{g}$ is time invariant.
Hence, with ${\bf e}_i(t)$ the appropriate evolution equations for 
magnetic field and density are
\begin{equation}\label{eq:inductgen}
\partial\tilde{B}^i/\partial t+\tilde{B}^j {\bf e}^i\cdot d{\bf e}_j/dt = \partial_3 \tilde{B}^i
\end{equation}
\begin{equation}\label{eq:mgen}
\partial \rho / \partial t = (1/\sqrt{g}) \partial_3 (\sqrt{g} \rho)
=\partial_3 \rho + \rho \partial_3 \log \sqrt{g}
\end{equation}
Thus in each \Eqs{inductgen}{mgen} there is an additional term which may cause growth,
unless constraints are placed on the metric specifically to make the term vanish.
The explicit time variation of the basis vectors appearing
in \Eq{inductgen} looks more easily avoidable,  since for \Eq{mgen}
the invariance of~$\sqrt{g}$ in a coordinate direction usually requires the
same invariance of the whole metric tensor and hence underlying mapping.
As the model incompressible equations lack the extra term,
there follows the implication that an incompressible, time
dependent flow is less likely to be a perfect dynamo than a
compressible time dependent flow.

\section{Examples of anti-dynamos}\label{sec:anti}
As should be evident from \Sec{math}, coordinate bases may be constructed
as a by-product of non-singular mappings of 3-D Cartesian space.
Ref~\cite[\S\,II.A]{Wa13c} explains that such mappings may be produced
by grid generation software. Incidentally, such software 
may ensure only minimal continuity requirements (in first derivative) on the basis vectors,
hence on the resulting anti-dynamo flows~${\bf u}$. Compare with the
lack of smoothness constraints on the initial ${\bf B}$-field, arising
because the pure advection equation supports discontinuous solutions.

The vanishing of the Lie bracket of the two solenoidal
vectors~${\bf B}$ and current~${\bf J}=\nabla\times{\bf B}$
is the condition for magnetic equilibrium (more familiarly $\nabla\times({\bf J}\times{\bf B})={\bf 0}$).
Thus, if the coordinate basis can be completed as in, say~\cite[\S\,III.D]{Wa13c},
the triad contains a basis vector~${\bf u}={\bf B}$ which not only has
the anti-dynamo property but is also the flow of a perfect fluid.

Magnetic equilibria in the solid torus are of particular interest for laboratory
magnetic confinement problems. Those of constant~$q$
are shown to yield coordinate bases in ref~\cite[\S\,VI.C]{Wa13c}.

\section*{Acknowledgement}\label{sec:ackn}
I am very grateful to Andrew Gilbert of Exeter University for
generously sharing with me his expertise in dynamo theory.
%This work, part-funded by the European Communities under the contract of
%Association between EURATOM and CCFE, was carried out within the framework
%of the European Fusion Development Agreement. The views and opinions
%expressed herein do not necessarily reflect those of the European
%Commission. This work was also part-funded by the RCUK Energy Programme
%under grant EP/I501045.
%This work was funded by the RCUK Energy Programme under grant EP/I501045 and
%the European Communities under the contract of Association between EURATOM and CCFE.
%The views and opinions expressed herein do not necessarily reflect those of
%the European Commission.
This work was funded by the RCUK Energy Programme grant number EP/I501045
and the European Communities under the contract of Association between
EURATOM and CCFE.
To obtain further information on the data and models underlying this
paper please contact PublicationsManager@ccfe.ac.uk.
The views and opinions expressed herein do not necessarily reflect
those of the European Commission.

%PRL These two lines to be removed
%\section*{References}
\bibliographystyle{unsrt}
%PRL

%\bibliography{waynes,misc,new,warv,neuts}
\end{document}